\documentclass[]{aa}
\usepackage{graphicx,txfonts}

\newcommand{\af}{Alfv\'{e}n }
\newcommand{\afw}{Alfv\'{e}n waves}

\newcommand{\Fg}[1]{Figure~\ref{#1}}

\newcommand{\kms}{km$\,$s$^{-1}$}

\begin{document}

\title{Large-amplitude transverse MHD waves prevailing in the H$\alpha$ chromosphere of  a  solar quiet region revealed by  MiHI integrated field spectral observations }
\titlerunning{Transverse MHD Waves Revealed by MiHI}
\authorrunning{Chae et al.}

\author{Jongchul Chae\inst{\ref{snu}}\thanks{Corresponding author: jcchae@snu.ac.kr} \and  Michiel van Noort\inst{\ref{mps}} \and Maria S. Madjarska\inst{\ref{mps},\ref{bul}} \and  Kyeore Lee\inst{\ref{snu}} \and Juhyung Kang\inst{\ref{snu}}
      \and  Kyuhyoun Cho\inst{\ref{baeri},\ref{lockh}} } 
\institute{Astronomy Program, Department of Physics and Astronomy, Seoul National University, Gwanak-gu, Seoul 08826, Korea \label{snu}
\and Max-Planck Institute for Solar System Research, Justus-von-Liebig-Weg 3, 37077 Göttingen, Germany \label{mps}
\and Space Research and Technology Institute, Bulgarian Academy of Sciences, Acad. Georgy Bonchev Str., Bl. 1, 1113, Sofia, Bulgaria \label{bul}
\and Bay Area Environmental Research Institute, NASA Research Park, Moffett Field, CA 94035, USA \label{baeri}
\and Lockheed Martin Solar \& Astrophysics Laboratory, 3251 Hanover Street, Palo Alto, CA 94304, USA \label{lockh}}

\abstract{
The investigation of plasma motions in the solar chromosphere is crucial for understanding the transport of mechanical energy from the interior of the Sun to the outer atmosphere and into interplanetary space. We report  the finding of large-amplitude oscillatory transverse motions   prevailing in the non-spicular H$\alpha$ chromosphere  of  a small quiet region near the solar disk center.   The observation was carried out on 2018 August  25 with the Microlensed Hyperspectral Imager (MiHI)  installed as an extension to the spectrograph at the Swedish Solar Telescope (SST).  MiHi produced  high-resolution Stokes spectra of the H$\alpha$ line over a two-dimensional array of points (sampled every 0.066\arcsec\  on the image plane) every 1.33 s for  about 17 min.  We extracted the Dopple-shift-insensitive  intensity data of the line core by applying a bisector fit to  Stoke I line profiles.  From our time--distance analysis of the intensity data, we find a variety of transverse motions with  velocity amplitudes of up to 40 \kms\  in fan fibrils and  tiny filaments.
In particular, in the fan fibrils,  large-amplitude transverse MHD  waves were seen to occur with  a mean velocity amplitude of 25 \kms\ and a mean period of 5.8 min, propagating at a speed of  40 \kms. These waves are nonlinear   and display group behavior.  We estimate the wave energy flux in the upper chromosphere at $3 \times 10^6$ erg~cm$^{-2}$~s$^{-1}$.
Our results  contribute to the advancement of our understanding of the properties of transverse MHD waves in the solar chromosphere.
}
\maketitle

\section{Introduction}

 Transverse magnetohydrodynamic (MHD) waves, such as \afw,\ have been regarded as one of the potential ways of transporting the mechanical energy  required  for coronal heating and solar wind  acceleration \citep[e.g.,][]{1947MNRAS.107..211A, 1971A&A....13..380A}.  An important point is that if  transverse waves are to be fully in charge of the coronal heating and the solar wind acceleration,  they  should have  large velocity amplitudes in the photosphere and chromosphere, because the transmission efficiency of  the waves in the solar atmosphere  is very low due to the rapid increase in \af speed with height \citep[e.g.,][]{1993A&A...270..304V}.  According to the recent theoretical study of \citet{2023ApJ...954...45C},   the velocity amplitude has to  be as large as 20 \kms\  in the upper chromosphere, both in  coronal holes and in active regions. This kind of theoretical consideration has motivated researchers to search for large-amplitude transverse waves in the chromosphere.

 The transverse waves detected in the chromosphere of active regions so far  are found to have velocity amplitudes of much smaller than 20 \kms.  These values were mostly determined from observations of superpenumbral fibrils around sunspots. The values  determined  from  transverse displacements  are typically 1 \kms\ \citep{2011ApJ...739...92P}, with a mean of 0.8 \kms\ \citep{2021RSPTA.37900183M}, and   those determined from the analysis of a  Doppler velocity pattern ranged from 0.6 to 1 \kms\ \citep{2021ApJ...914L..16C,2022ApJ...933..108C}.

In contrast, the transverse waves in quiet regions are found to have large velocity amplitudes.  The observations of spicules above the solar limb taken with the Solar Optical Telescope on board Hinode   revealed large-amplitude transverse oscillations of spicules with number distributions of velocity amplitude displaying tails extending up to 30 \kms\ \citep{2007Sci...318.1574D, 2011ApJ...736L..24O,2012ApJ...759...18P,2022ApJ...930..129B}.  From off-limb observations, it was also found that velocity amplitudes increase with height above the solar limb, with the mean value ranging from 18 \kms\  at  4900 km to 26 \kms\  at 7500 km \citep{2022ApJ...930..129B}.  Transverse waves were also reported from the disk counterpart of limb spicules, that is, fibrils around strong network elements.  \cite{2009Sci...323.1582J} found torsional oscillations with a velocity amplitude of 2.6 \kms\  at network bright points that may correspond to  spicule footpoints. From their observations of on-disk spicules, \cite{2012ApJ...744L...5J}  reported transverse velocities with a mean of  around 15 \kms\ and  reaching  up to 28 \kms.  \cite{2017NatSR...743147S} reported high-frequency transverse oscillations with velocity amplitudes of 14 \kms\   from the observation of  on-disk spicules. Observations of rapid blue excursions (RBEs), the supposed disk counterpart of Type II spicules,  revealed  transverse velocities with  a mean amplitude of around 12 \kms\ \citep{2013ApJ...764..164S},  reaching up to 22 \kms\ \citep{2015ApJ...802...26K}. All of these studies seem to support the notion that spicules carry transverse MHD waves with sufficiently large velocity amplitudes   for coronal heating in quiet Sun regions.

 We note, however, that spicules are not all   plasma structures  in the upper chromosphere of quiet regions. Spicules usually refer to plasma jets along the magnetic  field lines that emanate from strong network magnetic elements, and become predominantly vertical at large heights.  The disk counterparts of spicules were regarded as plasma structures that were called  mottles or bushes in the past, and are presently often referred to as fibrils. In the past, however, the term ``fibril'' was used to refer to plasma structures that are supported by highly inclined magnetic field lines either in active or quiet regions of the solar atmosphere \citep{1971SoPh...19...59F}.  Among nonspicular elongated plasma structures are also threads (or loops),  plasma structures connecting two magnetic poles of opposite polarity,  and filaments,  plasma structures located above the polarity inversion line.   It is generally considered that all spicules, mottles, fibrils, threads, and filament fine structures trace local field lines in the chromosphere \citep{1971SoPh...20..298F}, despite some exceptions reported from numerical modeling \citep{2016ApJ...831L...1M} or from  spectropolarimetric observations \citep{2017A&A...599A.133A}.

  Observations of transverse waves in the nonspicular chromosphere of the quiet Sun are scarce. \cite{2012NatCo...3.1315M} and \cite{2013ApJ...768...17M} are among the few authors to have reported such rare observations.
 They  found  transverse oscillations with mean velocity amplitudes of 6.4 \kms\ and 4.5 \kms\ in nonspicular fibrils, respectively.  Very recently, \cite{2023ApJ...958..131K} reported  transverse oscillations in the line-of-light direction with  a mean velocity amplitude of 1.3 \kms\ in quiet region fibrils  that would be better referred to as chromospheric threads or  loops. 

 In this paper, we report results from observations carried out with an integral field spectrograph covering a very small quiet region on the solar disk with a nearly diffraction limited resolution (0.12\arcsec), a high cadence (1.33~s), and a spectral resolution of R>300000, covering a wavelength range of 4.5~\AA\ ($\pm 100$ km\,s$^{-1}$).
The H$\alpha$ chromosphere of the observed region happened to contain a number of   fibrils and tiny filaments   supported by predominantly horizontal magnetic fields. We found  a variety of oscillatory transverse motions with mean velocity amplitudes of 9 \kms, 25 \kms,  and 37 \kms\ in different groups.  This is the first report on large-amplitude transverse oscillations prevailing in the nonspicular chromosphere of the quiet Sun. 



\section{Data and analysis}
We used the H$\alpha$ spectral data  taken  from a quiet region of the Sun  on 2018 August 25 with the Microlensed Hyperspectral Imager (MiHI) prototype \citep{2022A&A...668A.149V,2022A&A...668A.150V}. The MiHI is an integral field spectrograph based on a double-sided microlens array (MLA) that was installed as an extension to the TRI-Port Polarimetric Echelle-Littrow (TRIPPEL)  spectrograph at the Swedish Solar Telescope (SST).
 For information on the observed region, we  use a photospheric magnetogram  taken by the Heliospheric and Magnetic Imager (HMI) and images taken by the Atmospheric Imaging Assembly (AIA) on board the Solar Dynamics Observatory (SDO).  The subregion of the HMI magnetograms that has a field of view (FOV) much larger than that of the MiHI was used to calculate potential magnetic fields in the chromosphere and corona.
 The AIA images sample emission at coronal (AIA 171~\AA), transition region (AIA~304~\AA), and chromospheric (AIA~1700~\AA) temperatures.

  The MiHI observations started at  09:07:48 UT and lasted for  1037 s, with a cadence of  1.33 s per frame.  The FOV was  $9.2\arcsec \times 8.2\arcsec$   (6700 km by 6000 km)   and the spatial sampling was 0.066\arcsec\ (48 km). The conversion of the raw data frames to hyperspectral cubes and their restoration to high-resolution science-ready Stokes data was done following the data reduction procedure described by \cite{2022A&A...668A.151V}.

 Every pixel contains the spectral profile covering  3.4 \AA\ around the H$\alpha$ line with sampling of 0.01\AA.  We note that the spectral data were taken simultaneously at all the pixels without any time difference, as the MiHI is  an integral field spectrograph. The data reduction process consists of  bad-pixel mending, wavelength recalibration, telluric-line removal, and noise suppression based on the principal component analysis. The line intensity at every wavelength is expressed in units of  continuum intensity, which is assumed to be equal to 1.18 times  the offband intensity at $-1.5$ \AA\   averaged over the FOV and over the observing time.

 We obtained the core intensity and Doppler shift of the H$\alpha$ line by applying a  0.5 \AA\ width bisector  fit to every line profile.  The Doppler velocities  are found to have a significantly large variation with a standard deviation of 4.1 \kms.
 In principle, the combination of plane-of-sky transverse motion and line-of-sight Doppler motion  should provide more information on oscillations and waves.  However, the  pattern of Doppler motion is found to be too complex for straightforward analysis.  Moreover, in order  to  infer the line-of-sight motion of a feature that is overlapped by other structures along the line of sight, a  sophisticated spectral analysis other than the simple bisector method  is required, which is far beyond the scope of this work.

   In the present work, we focus on the inference  of transverse motion from the core intensity images. We note that the core intensity data  were obtained from the spectral analysis, and  are therefore insensitive to Doppler shifts.  The core intensity is rather related to the H$\alpha$ line source function $S$ and optical thickness $\tau$  of the observed feature. We note that the H$\alpha$ line source function $S$ is  constant over the wavelength to a good approximation.  It is also commonly assumed that $S$ is constant over the position along the  line of sight.  The equation of radiative transfer along the line of sight can then be solved to yield the expression
\begin{equation}
\Delta I  \equiv I - I_0 = (S - I_0) (1- \exp(-\tau) )
,\end{equation}
 given in terms of background core intensity $I_0$ and  optical thickness $\tau$  as well as $S$.
 This expression is reduced to $\Delta I = S-I_0$ in the limit $\tau \gg 1$, and to $\Delta I = (S-I_0) \tau$ in the limit $\tau \ll 1$. We note that all  $S$,  $I_0,$ and $\tau$  are now functions of time and position on the image plane.  In this study, as an approximation, we identify $I_0$ with the slow-varying pattern that is constructed from the low-order polynomial fit of the temporal variation of the intensity at every position.

\section{Results}

\begin{figure}[t]
\centerline{\includegraphics[width=8.5cm]{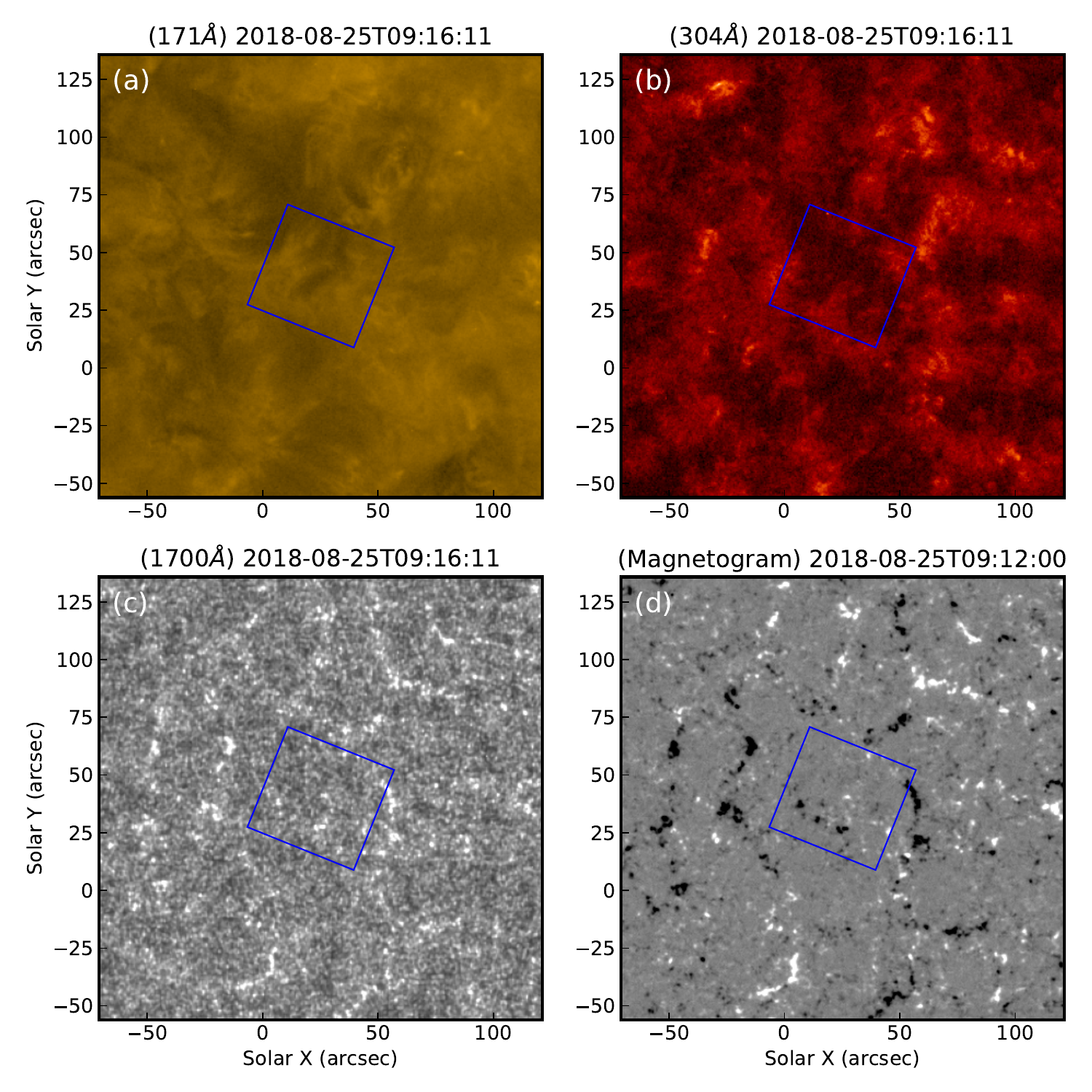}}
\centerline{\includegraphics[width=8.5cm]{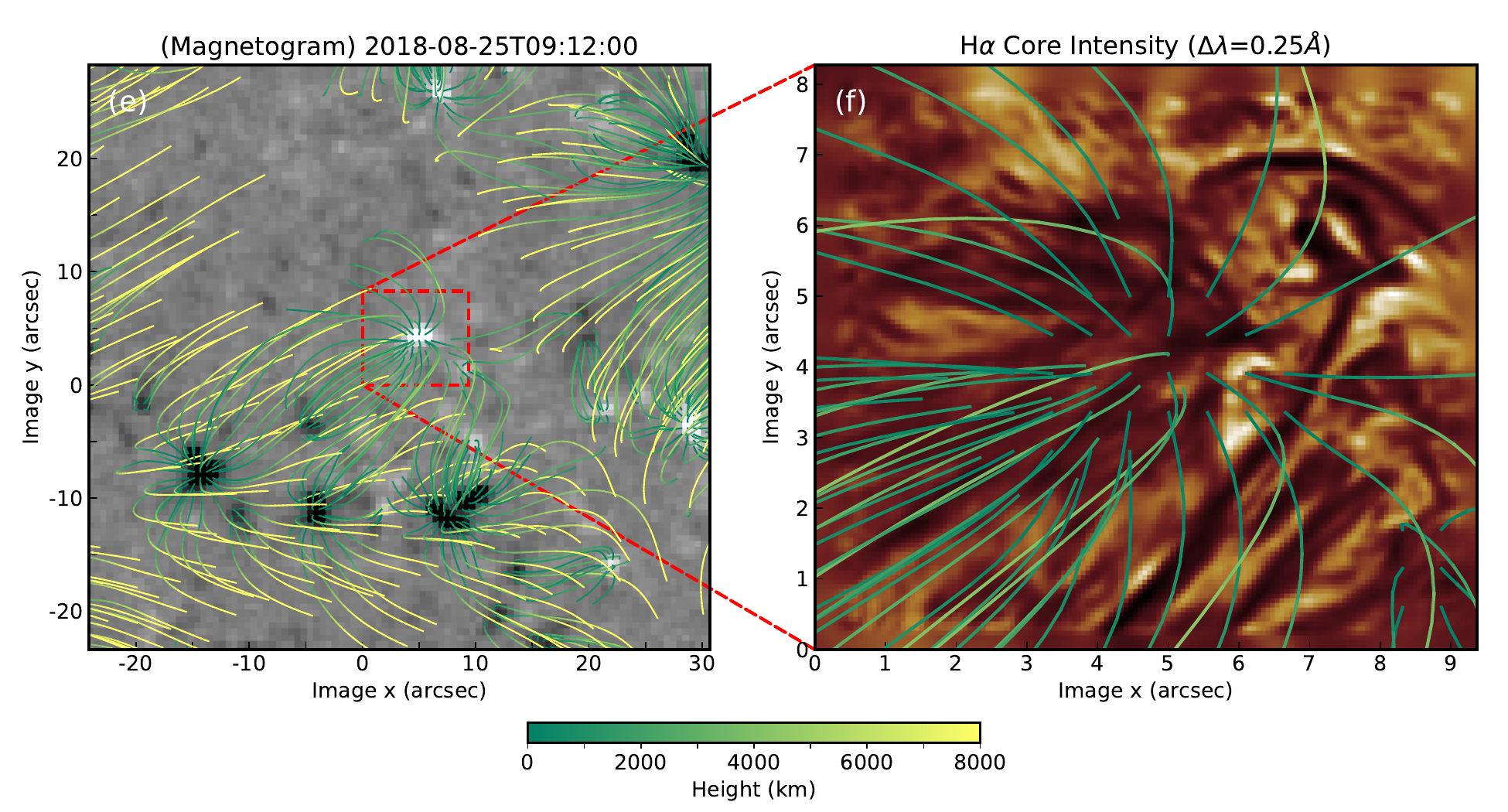}}
\caption{ Images of the observed region. \textsf{a, b, c}:
Context SDO/AIA  images. \emph{d}: SDO/HMI magnetogram. The blue rectangle indicates the 55\arcsec $\times$51\arcsec\  FOV of the  submagnetogram below.  \textsf{e}:  SDO/HMI submagnetogram rotated to match the MiHI image frame where the origin is located in the lower left corner of  the  9.2\arcsec $\times$8.2\arcsec\ MiHI FOV (red rectangle).  The colored curves denote the projected magnetic field lines of the constructed potential  field configuration with color representing height  above the surface.  \textsf{f}: MiHI H$\alpha$ intensity image with  the projected field lines superimposed.   Most of these field  lines inside the MiHI FOV have heights of $<$ 4000 km.   \label{fg:FOV1}}
\end{figure}

\begin{figure}[t]
\centerline{\includegraphics[width=8cm]{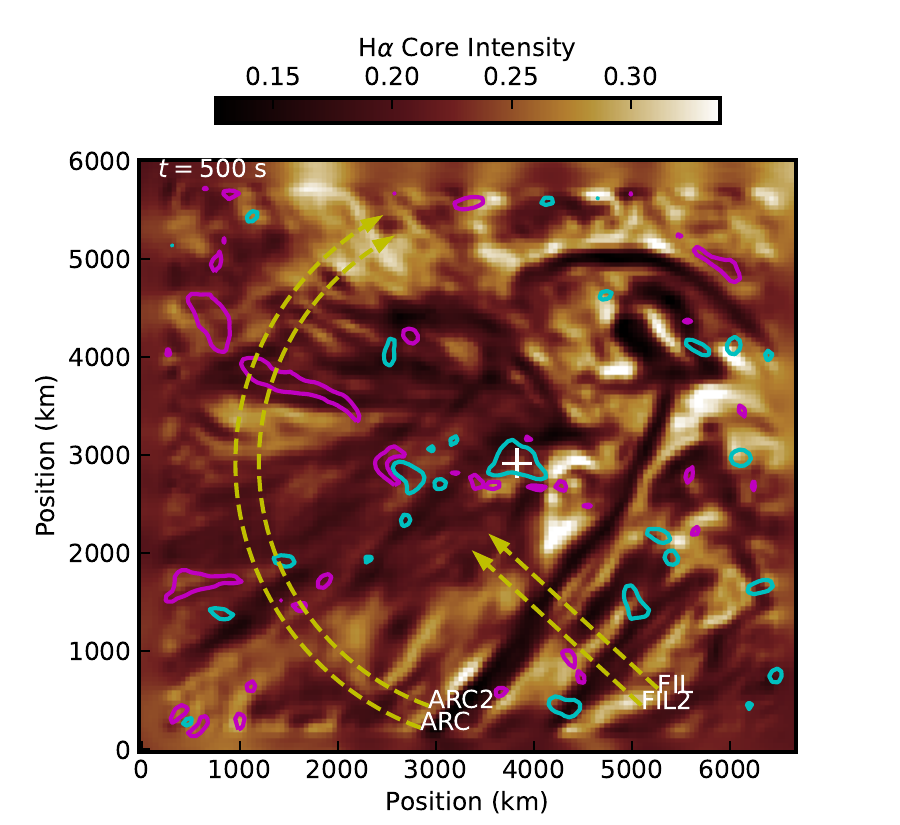}}
\caption{H$\alpha$ core intensity map taken at 500 s after the start of observations.  The dashed arrows indicate  the two main slits (FIL and ARC) and the two auxiliary slits (FIL2 and ARC2).  The cyan-colored and magenta-colored contours  represent $\pm 1\sigma$ levels of offband intensity at $-1.5$ \AA, and  the plus symbol marks the center of the offband bright concentration co-spatial with the magnetic concentration of positive polarity.  \label{fg:FOV}}
\end{figure}

\Fg{fg:FOV1} shows the quiet region observed by MiHI  as  marked on  SDO/AIA images and SDO/HMI  magnetogram. This region does not show any noticeable bright coronal feature  as seen in these images apart from faint small-scale loops. The HMI  magnetogram indicates that the MiHI FOV contains one magnetic field concentration of positive polarity with a flux density of up to 90 G. We therefore conclude that the observed region was very quiet in the
extreme ultraviolet (EUV)) and magnetograph observations during  the MiHI observation  and may be typical of quiet regions on the Sun. This implies that any process we observe in this FOV may not be particular to this region, but  rather commonly occurring in other regions of the quiet Sun as well.

\Fg{fg:FOV} illustrates the H$\alpha$ core intensity image of the observed region taken at one particular instant.   The figure shows a number of elongated absorption structures.
We find from the time series of the core intensity images  that  these structures change ceaselessly, even over just 17 min. A variety of motions occur in this small region in such a short time, including transverse motions perpendicular to the length, longitudinal motions, and large-scale rotating motions leading to the reorganization of the structures themselves.
  Here we focus on  the time--distance ($t$-$d$) analysis of transverse motions perpendicular to the length.

   There are two categories of  absorption structures in our observation: fan fibrils and tiny filaments.
  The fan fibrils are  lower-contrast features comprising a fan structure  extending approximately radially outward from  the  positive magnetic concentration inside the MiHI FOV.  These fibrils are likely to be supported by highly horizontal, low-lying magnetic field lines connecting this magnetic concentration and some magnetic concentrations of negative polarity outside the MiHI FOV.   The constructed potential magnetic fields demonstrate the existence of these field lines.
   The potential field lines  are fairly horizontal, and are mostly located at heights  of 2000 to 4000 km  \ (see Figure~\ref{fg:FOV1}f).   The  field strength at the loop tops at these heights is estimated  at around 10 G.
   It is not surprising to find some misalignments  between the fibrils and the potential field lines, because the effects of plasma pressure and gravitational force may not be negligible in the chromosphere.  If these effects were taken into account,  the field lines would become  flatter, and  the magnetic field strength would then be  a little higher than 10 G.
   For the study of the transverse motions  in these fan fibrils, we define the arc-shaped slits ARC and ARC2  for the  $t$-$d$ analysis, as shown in Figure~\ref{fg:FOV1}.

  The tiny filaments are higher-contrast absorption structures outside the fan structure.  These structures do not originate from the magnetic concentration, and are not aligned  with the constructed potential fields   (see Figure~\ref{fg:FOV1}\textsf{f}).
 It seems that these features  are supported by highly nonpotential and predominantly horizontal magnetic fields.   This is one reason why we identify them as   filaments.
For the study of transverse motion in these  filaments,  we define the two straight  slits FIL and FIL2  for  the $t$-$d$ analysis.

 We find  a variety of large-amplitude velocity oscillations in this region.  We categorize  the transverse oscillatory motions into three groups; those with velocity amplitudes $V< 15$ \kms  (Group I),  those with  $20 < V <  30 $ \kms\  (Group II),  and  those with  $V> 30$ \kms\ (Group III). Group I  oscillations  occurred continually in the filaments,   and the Group II oscillations prevailed in the fan fibrils for about 350 s among the observing duration of 1037 s.  Group III oscillations occurred from time to time in the fan fibrils, less  frequently than the oscillations in the other groups.

\begin{figure*}[ht]
\centerline{\includegraphics[width=14cm]{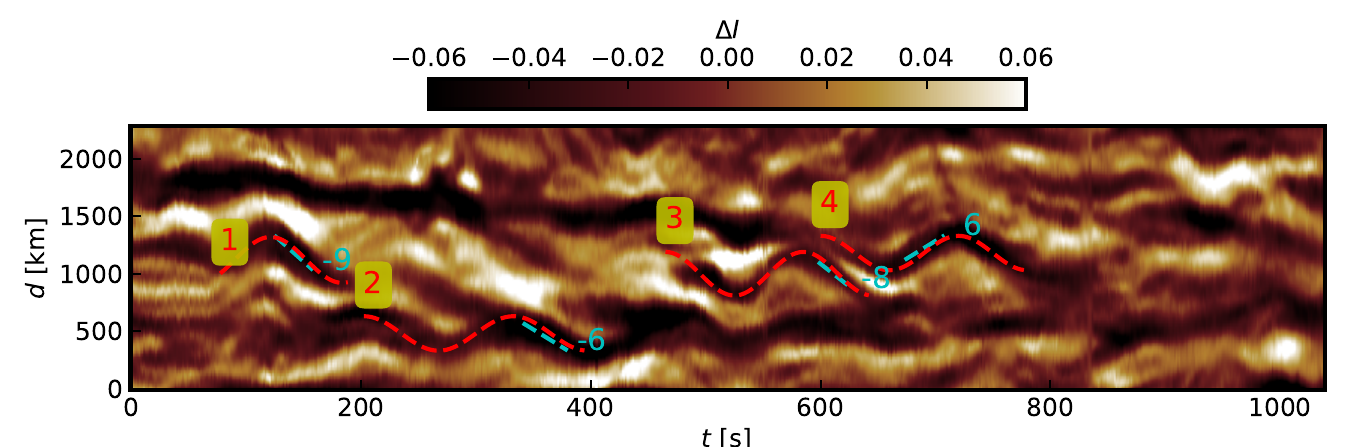}} \caption{Time--distance  $\Delta I$ map constructed along the slit  FIL.   Each cyan-colored dashed line segment  traces the trajectory of a filament at each instant, and its slope corresponds to the instantaneous velocity at one particular instant, as  explicitly specified by a cyan-colored number (in units of \kms). Each red sinusoidal curve traces the transverse oscillation of the  filament,  which is indexed by  red numbers ranging from 1 to 4.    \label{fg:FIL} }
\end{figure*}

 Transverse oscillations belonging to Group I occur mostly in the filaments, and are easily identified from  the $t$-$d$ map of intensity constructed along the slit FIL shown in  \Fg{fg:FIL}. These fibrils are characterized by contrast  $\Delta I \sim  -0.06$ and a full width at half maximum (FWHM) of $w \sim  200 $ km (see Table~\ref{tb:par}).
In \Fg{fg:ARC}, we  mark some line segments representing our manual tracing of some noticeable displacements of the filaments.  The slope of each line segment yields the instantaneous velocity at a particular instant; these range from 6 to 9 \kms.  These instantaneous motions are parts of an oscillatory motion, and the velocity amplitude of the oscillatory motion is  larger  than or equal to any measured instantaneous speed.
 To determine the velocity amplitude and period  of each oscillatory motion, we   manually traced the oscillatory motion  by drawing on the $t$-$d$ map  a sinusoidal curve of the form $d = D \sin (2 \pi (t-t_0)/P)$    that best matches the pattern of the oscillatory displacement.
As a result, we  found oscillations (FIL:1,2,3, 4) with periods $P$ of 120 -- 130 s,  and  velocity amplitudes of $V \equiv 2 \pi D/P$ of 7 --11 \kms\ as listed in Table~\ref{tb:par}.  An oscillation in Group I   was also identified in one of the fan fibrils from
  the $t$-$d$ map constructed along the slit ARC shown in \Fg{fg:ARC}.
  This oscillation (ARC: 1) has a velocity amplitude of 11 \kms, which is comparable to those of FIL: 1-- 4,
   and has a period of 260 s, which is longer than those seen at the slit FIL.

\begin{figure*}[ht]
\centerline{\includegraphics[width=14cm]{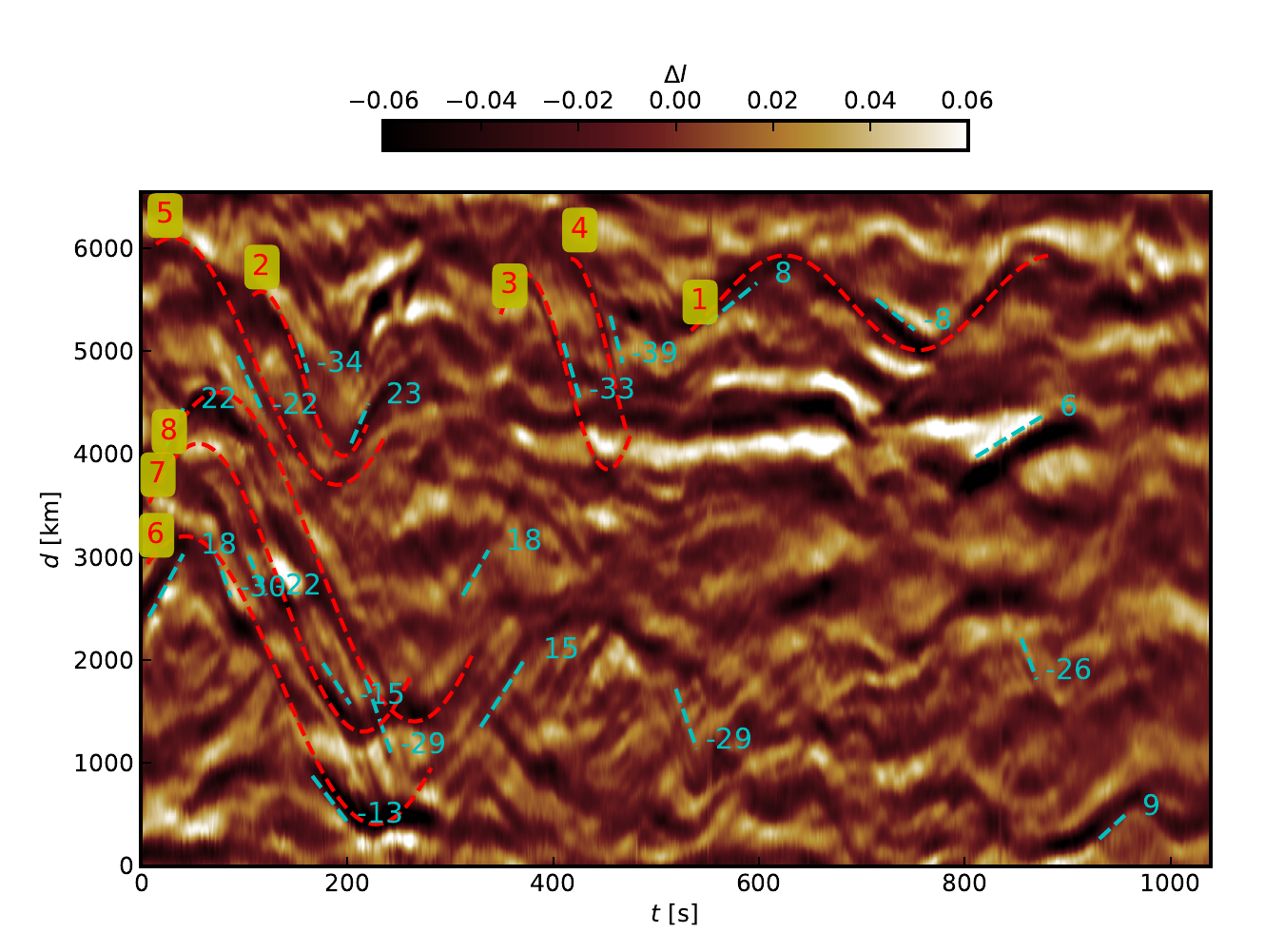}}
\caption{Time--distance $\Delta I$ map  constructed  along
the slit ARC.  Each cyan-colored dashed line segment  traces the trajectory of a fibril at each instant, and its slope corresponds to the instantaneous velocity at one particular instant, as  explicitly specified by a cyan-colored number (in unit of \kms). Each red sinusoidal curve traces the transverse oscillation of a fibril, which is indexed by  red numbers ranging from 1 to 8.
\label{fg:ARC} }
\end{figure*}

\begin{table}
\caption{Inferred values of  the period $P$ and velocity amplitude $V$  of the transverse oscillations, and the intensity difference $\Delta I$ (in units of continuum intensity), and the FWHM width $w$ of the fibrils  and filaments.   \label{tb:par}}

\centerline{
\begin{tabular}{c|c|cc|ccc}
\hline \hline
  Group & Slit: No  & $P$    &  $V$   &  $\Delta I$  &  $w$   \\
   &          &   s      &  \kms          &    &    km           \\ \hline
 I & FIL: 1     &      125  &  10    &     -0.05     &  210  \\
 & FIL: 2          &    130   &  7   &        -0.07   &  205   \\
 & FIL: 3       &    120   &  10   &      -0.04      &  160   \\
 &FIL : 4        &   120     &  8     &      -0.06    &   195 \\
  & ARC: 1       &  260      &  11   &      -0.05     &   160    \\  \hline
  II & ARC: 5   &  320  &   24  &  - &  -  \\
   & ARC:  6   &  370   &  24    &   - & -   \\
 &  ARC: 7 &    320   & 27  & -  & - \\
 &  ARC:  8  & 380 & 26 &  - & -   \\ \hline
  III & ARC: 2 & 160  &  31  & -0.03 & 130  \\
  & ARC: 3 &  160 &  37  &  -0.02 & 140  \\
       &   ARC: 4 &  160 & 43  &  -0.02   &  145 \\
      \hline
\end{tabular}}
\end{table}

   \Fg{fg:ARC} shows the transverse oscillations in Groups II and III as well.   The transverse oscillations (ARC: 5 to 8 )  in Group II are distinct from those in Group I in several ways.  First, unlike those in Group I,  the trajectories of the Group II oscillations are not  uniform, either in intensity or in width, and appear fragmented at certain instances.  This kind of trajectory fragmentation might be because short fibrils move in and out of the slit  along the fibril length.  We  manually drew the sinusoidal curves  to approximate the fragmented trajectories as shown in the figure.  We believe that these curves, though imprecise, reasonably characterize the oscillatory transverse motions.  The probable errors  in the determination of $D$ are responsible for  most of the probable errors in $V$,  which are estimated to  be about 2 \kms\ in three oscillations (ARC: 5, 6, 8) and  about 8 \kms\ in the other oscillation (ARC: 7).    We find that these oscillations are characterized by larger velocity amplitudes  (from 24 to 27 \kms) and longer periods (320 to 380 s)  than those in Group I (see Table~\ref{tb:par}).

    An important property of the transverse oscillations in Group II  (ARC: 5 to 8) is the group behavior. These oscillations are not independent of one another; they have similar phases,  periods, and amplitudes, even though they occurred at different locations on the cut ARC.  The consistency in these four oscillations  strongly suggests that they may have originated from the same source. 

 The transverse oscillations (ARC: 2 to 4) in Group III are characterized by very large velocity amplitudes, above 30 \kms, and short periods of about 160 s.  The trajectories are regular, but the identified durations  are much shorter than the period. The fibrils displaying these oscillations  are less
 dark ($\Delta I \sim  -0.02$) and are thinner ($w \sim  140 $ km) than those in Group I.  These fibrils seem to appear at higher altitudes.  It is noteworthy that the periods and widths of the quiet Sun features summarized in Table 1 are comparable with the reported lifetimes and widths of dynamic fibrils in active regions \citep[e.g.,][]{2012SSRv..169..181T}.

 Finally,   in addition to the transverse oscillations, the $t$-$d$  map in \Fg{fg:ARC} displays a number of short-lived (< 60 s) transverse motions as inferred from the faint intensity streaks (marked by dashed line segments). Their slopes correspond to speeds from 6 to 29 \kms. As these speeds are in the same range as the oscillatory motions examined above,
these streaks are probably of the same kind as the oscillatory motions.

Now we attempt to determine the propagation speed of the transverse oscillations.  For the transverse oscillations in Group I, we constructed another time--distance map  along the slit FIL2, which is parallel to FIL and is 240 km away from it (lower left from FIL).  The  time lags obtained from the transverse motions of FIL: 1, 3, and  4 are about 4  s. This means that the oscillations propagate along the filaments at a speed of about 60 \kms.
 For the oscillations in fan fibrils, we constructed another $t$-$d$ map  along the slit ARC2, which is parallel to ARC but has a smaller radius, with the distance between the two slits  being 240 km.  From the transverse oscillation ARC: 6 in Group II,   we roughly estimate the time lag at about +6 s,  which corresponds to  a speed of about 40 \kms.  For the transverse oscillations in Group III,  we obtained a time lag of about -5 s,  corresponding to a velocity of $-48$ \kms. The inward waves implied by the negative velocity might have originated either from wave excitation  at a remote magnetic concentration of negative polarity or  from  wave reflection at a place  where the gradient of the \af\ speed is large  \citep[e.g.,][]{1993A&A...270..304V}. The propagation speeds inferred here are quite compatible with the theoretical range of  \af\ speed,   15 to 91 \kms,  in the upper chromosphere of the quiet Sun obtained with the field strength of 10 G, and  the  range of mass density from  $3.5 \times  10^{-12}$ g cm$^{-3}$  to $9.5 \times  10^{-14}$ g cm$^{-3}$ taken from an atmospheric model of the quiet Sun \citep{1993ApJ...406..319F}.   This supports the idea that the observed transverse oscillations represent  transverse MHD waves propagating at speeds comparable to the \af speed in the upper chromosphere.

\section{Discussion}

  In the present work we report the observation of a small region on the Sun that appears to be highly typical of the quiet Sun.  The H$\alpha$ chromosphere of this region contains two kinds of  elongated absorption structures:  fan fibrils supported by highly horizontal magnetic fields and tiny filaments  supported by nonpotential magnetic fields.
The H$\alpha$ spectral observations of high angular resolution
and high temporal resolution obtained with the MiHI prototype  reveal that this region was very dynamic, displaying a variety of transverse motions with velocity amplitudes of up to 40 \kms.    Interestingly, however, no noticeable activities were
seen in the EUV image data.

 Particularly important is our finding of large-amplitude transverse oscillations in fan fibrils (Group II oscillations). The velocity amplitudes of these oscillations  have
 a mean value of  about 25 \kms.
   We note that such mean velocity amplitudes were previously only reported from observations of spicules seen high above the limb; for example,  7500 km above the limb \citep{2022ApJ...930..129B}.
   Our study is the
first to report a large mean value of velocity
amplitudes obtained from on-disk observations of the nonspicular chromosphere at low heights $<$ 4000 km.

 We now consider our estimates of the wave energy flux in the chromosphere and corona.
We  estimate the \af\ wave energy flux   in the chromosphere at  $2.8 \times 10^6$ erg cm$^{-2}$ s$^{-1}$ from  the observed mean velocity amplitude of 25 \kms\  and the magnetic field strength of 10 G;  for these estimates, we adopted a mass density  of  $1.0 \times 10^{-13}$ g cm$^{-3}$.   Furthermore, by multiplying the theoretical transmission of 0.36  \citep{2023ApJ...954...45C}, we estimate the wave energy flux in the corona  at $9.5 \times 10^5$ erg cm$^{-2}$ s$^{-1}$.  This estimate is comparable to the wave flux of $8.0  \times  10^5$ erg cm$^{-2}$ s$^{-1}$ required for coronal heating and solar wind acceleration in corona hole regions \citep{2023ApJ...954...45C}.
 Contrary to our expectations, we did not detect any noticeable activity in the coronal image data, despite the occurrence of the large-amplitude transverse waves in the fan fibrils. This could be explained by the fact that the guiding magnetic field lines are strongly inclined to the horizontal, forming low-lying magnetic loops that are not directly connected to the corona above, unlike the spicular chromosphere. Nevertheless, the transverse waves may still contribute to the heating of these low-lying magnetic loops. Moreover, it would be natural to suppose that large-amplitude transverse waves as observed in the fan fibrils take place also in the spicular chromosphere, contributing much to the coronal heating and solar wind acceleration.

 We note that the transverse MHD waves we observed    not only have large velocity amplitudes, but are also significantly {nonlinear}. The nonlinearity of these waves  is indicated by the large value of the observed mean velocity amplitude  25 \kms\ in comparison with the estimated propagation speed of 40 \kms.  It is well known from theoretical studies \citep{1982SoPh...75...35H, 1999ApJ...514..493K} that  nonlinear \afw\ are coupled with compressible waves; these \afw\
 excite longitudinal motions that propagate as slow or fast waves and develop into shock waves. This kind of nonlinear coupling seems to be  consistent with  the properties of the oscillations  we observed in Group II.  The variations of intensity and width in the trajectories may be attributed to the co-existence with compressible waves as in the previous studies \citep{2012ApJ...744L...5J,2012NatCo...3.1315M,2018ApJ...853...61S},   and  the fragmented appearance may be due to the presence of oscillatory longitudinal motions in and out of the slit.

 It is also interesting that the transverse oscillations in Group II display group behavior. The transverse oscillations occurring in different fibrils have similar phases, velocity amplitudes (typically 25 \kms), and periods (5--7 min). This group behavior may provide us with a clue as to the excitation process of the waves.  These oscillations may have originated from the magnetic concentration where the magnetic field lines supporting the fibrils are rooted together.  If the magnetic concentration is disturbed either  by a linear motion or a twisting motion near the photosphere or below it,  transverse velocity oscillations will be produced.  These oscillations propagate  simultaneously along different field lines, developing into transverse oscillations of different fibrils, and displaying the group behavior.
In this regard, the observed  group behavior is not strange, and may be a very natural property of fan fibrils.  This property is not limited to fibrils of this kind, but may also appear in the spicules that emanate from the same magnetic concentration,  as previously identified as torsional \afw\  \citep{2017NatSR...743147S}.

 In summary, our results  indicate that large-amplitude transverse waves occur not only in the spicular chromosphere at large heights but also in the  nonspicular chromosphere at small heights.  The interesting properties (nonlinearity and group behavior)  of the transverse waves in fan fibrils  merit further systematic investigation  based on  the rigorous analysis of Doppler velocity data.  Together with the previously reported results on transverse waves in the spicular chromosphere, our results contribute to the advancement of our understanding of the properties and excitation of  transverse MHD waves  in the H$\alpha$ chromosphere of the quiet Sun.

\begin{acknowledgements}
 We appreciate the referee's critical comments that helped in our  assessment of the scientific  significance of this work.  J.C. is grateful to Sami Solanki for the travel support  and Hans-Peter Doerr for the hospitality provided during his visit to MPS in the summer of 2023.
This work was supported by  the National Research Foundation of Korea ( RS-2023-00208117) and (RS-2023-00273679).  M.M. acknowledges DFG grants WI 3211/8-1 and 3211/8-2, project number 452856778.  The Swedish 1-m Solar Telescope is operated on the island of La Palma by the Institute for Solar Physics of Stockholm University in the Spanish Observatorio del Roque de los Muchachos of the Instituto de Astrofísica de Canarias. The Institute for Solar Physics is supported by a grant for research infrastructures of national importance from the Swedish Research Council (registration number 2021-00169).
\end{acknowledgements}

\bibliographystyle{aa}

\end{document}